\documentclass[namedreferences]{SolarPhysics}
\usepackage[optionalrh,solaenum]{spr-sola-addons} % For Solar Physics 
\usepackage{graphicx}                    % For eps figures, newer & more powerfull
\usepackage{color}                       % For color text: \color command
\usepackage{url}                         % For breaking URLs easily trough lines
\begin{document}

\begin{article}

\begin{opening}

\title{A Curious History of Sunspot Penumbrae}

%%%%%%%%%%%%%%%%%%%%%%%%%%%%%%%%%%%%%%%%%%%%%%%%%%%
%% Authors Names
%
\author{D.~H.~\surname{Hathaway}$^{1}$}

%%%%%%%%%%%%%%%%%%%%%%%%%%%%%%%%%%%%%%%%%%%%%%%%%%%
%% Runningheads
%
\runningauthor{Hathaway}
\runningtitle{A Curious History of Sunspot Penumbrae}

%%%%%%%%%%%%%%%%%%%%%%%%%%%%%%%%%%%%%%%%%%%%%%%%%%%
%% Affilations 
%
  \institute{$^{1}$ NASA Marshall Space Flight Center, Huntsville, AL 35812 USA
                     email: \url{david.hathaway@nasa.gov}
             }

%%%%%%%%%%%%%%%%%%%%%%%%%%%%%%%%%%%%%%%%%%%%%%%%%%%
%%% Abstract 
\begin{abstract}
Daily records of sunspot group areas compiled by the Royal Observatory, Greenwich, from May of 1874 through 1976 indicate a curious history for the penumbral areas of the smaller sunspot groups. On average, the ratio of penumbral area to umbral area in a sunspot group increases from 5 to 6 as the total sunspot group area increases from 100 to 2000 $\mu$Hem (a $\mu$Hem is $10^{-6}$ the area of a solar hemisphere). This relationship does not vary substantially with sunspot group latitude or with the phase of the sunspot cycle. However, for the sunspot groups with total areas $<100 \mu$Hem, this ratio changes dramatically and systematically through this historical record. The ratio for these smallest sunspots is near 5.5 from 1874 to 1900. After a rapid rise to more than 7 in 1905 it drops smoothly to less than 3 by 1930 and then rises smoothly back to more than 7 in 1961. It then returns to near 5.5 from 1965 to 1976. The smooth variation from 1905 to 1961 shows no indication of any step-like changes that might be attributed to changes in equipment or personnel. The overall level of solar activity was increasing monotonically during this time period when the penumbra-to-umbra area ratio dropped to less than half its peak value and then returned. If this history can be confirmed by other observations ({\it e.g.} Mt. Wilson or Kodaikanal) it may impact our understanding of penumbra formation, our dynamo models, and our estimates of historical changes in the solar irradiance.
\end{abstract}

%%%%%%%%%%%%%%%%%%%%%%%%%%%%%%%%%%%%%%%%%%%%%%%%%%%
%% Keywords
%

\end{opening}
%-------------------------------------------------

%%%%%%%%%%%%%%%%%%%%%%%%%%%%%%%%%%%%%%%%%%%%%%%%%%%
%% Sections
%
\section{Introduction} 

While the earliest telescopic observations of the Sun did reveal the umbral-penumbral structure of sunspots, the physical nature of sunspots remained a complete mystery until Hale's discovery of intense magnetic fields in sunspot umbrae \cite{Hale08}.
Numerical models of sunspots now reveal how magnetic fields govern the structure of sunspot umbrae and penumbrae \cite{RempelSchlichenmaier11, BorreroIchimoto11}.
Penumbrae are now recognized as regions where the largely vertical fields in the umbrae spread out to become more horizontal.
These strong horizontal fields then alter the convective motions to produce the penumbral filaments and flows.
While current magnetohydrodynamical (MHD) models do produce penumbral filaments and associated flows, they have not fully addressed the question of why penumbrae are as big as they are.
They do, however, suggest that penumbral size and structure are influenced by the magnetic fields from nearby sunspots \cite{Rempel_etal09}.
 
The earliest study of the relative sizes of umbrae and penumbrae suggested a fairly constant ratio.
\inlinecite{Nicholson33} used Royal Observatory, Greenwich (RGO) data from 1917 to 1920 in his study of magnetic field intensity as a function of umbral area and noted that the ratio of the area of the penumbra to the area of the umbra is, on average, $\approx4.7$ for unipolar sunspots or the preceding members of bipolar sunspots.

\inlinecite{Waldmeier39B} measured umbral diameters and penumbral diameters of 53 sunspots photographed by Wolfer at Z\"urich in the years 1896, 1897, 1907, and 1917.
He examined the ratio of the penumbral to umbral diameters as a function of penumbral diameter and found an average value consistent with the ratio of areas given by \inlinecite{Nicholson33} but with a penumbra-to-umbra area ratio that decreased from 6.8 to 3.4 as the sunspot size increased from 100 $\mu$Hem (a $\mu$Hem is $10^{-6}$ the area of a solar hemisphere)to 1000 $\mu$Hem.

\inlinecite{Jensen_etal55} studied 653 single regular sunspots with areas $> 50 \mu$Hem found in the RGO {\it Photoheliographic Results} for the years 1878-1945.
They found that during times of sunspot cycle maxima the ratio of penumbral area to umbral area was a decreasing function of sunspot size -- but with a much weaker variation than was found by Waldmeier with his much smaller sample.
During times of sunspot cycle minima the ratio was, in general, lower with even less variation with sunspot size.
In a follow-on study \inlinecite{Jensen_etal56} included data up to 1954, again limited their data to that for 845 single sunspots, and found similar results.

\inlinecite{TandbergHanssen56} extended the work of \inlinecite{Jensen_etal55} to double and composite sunspots in the RGO data and found that the ratio of penumbral-to-umbral area was higher for the more complex sunspot groups.
When all sunspots were included, he found that this ratio was fairly independent of size for large sunspots but increased with sunspot size for the smaller sunspots.
\inlinecite{Antalova71} also found this increase in a study of 12,532 sunspot groups in the RGO data.

While the earlier studies on single sunspots indicated that the relative penumbral area decreased for larger sunspots, these last two studies indicated that large sunspot groups (with multiple sunspots) tend to have larger penumbral areas.

The relative sizes of penumbrae and umbrae are thus related to the surrounding field conditions ({\it e.g.} horizontal magnetic fields from nearby sunspots and/or plage).
Changes in the ratio of the penumbral and umbral areas would indicate changes in sunspot group structure.
This would have implications for both the Sun's magnetic dynamo and consequences for solar irradiance variations.

Recently \inlinecite{PennLivingston06} reported measuring a temporal decrease in the average umbral magnetic field strength accompanied by a temporal increase in the average umbral brightness of sunspots since the maximum of cycle 23.
If these results are not the consequence of selecting sunspots of different sizes at different times, then the simplest explanation would be that there were relatively fewer small sunspots during the maximum of cycle 23 - suggesting changes in the solar dynamo.

Reconstructions of past irradiance variations \cite{FoukalLean90} use total sunspot areas and assume that the ratio of penumbra-to-umbra area is unchanged.
If there are significant changes to the sunspot population, conclusions drawn from these reconstructions are compromised.

Here we examine the ratio of penumbral-to-umbral areas and find results consistent with the earlier work but with indications of curious long-term variations which appear to be unrelated to the level of solar activity.
If these variations can be confirmed by other data sources then there may be consequences for solar dynamo theory, for solar irradiance reconstructions, and for theories of penumbra formation.

\section{Sunspot Area Data}

Many observatories and individual observers have recorded visual observations of sunspots.
These records can be as simple as noting the number of sunspot groups and/or sunspots on the Sun, or as complex as detailed measurements of sizes and positions from photographs and/or drawings.
Of the many different observatories, the RGO stands out for the length of its record (1874 through 1976), the completeness of its record (99.7 \%), and the details of the record (daily positions of sunspot groups with both umbral and whole sunspot areas).
While extensive photographic collections do exist for other observatories ({\it e.g.} Mt. Wilson from 1917 and Kodaikanal from 1904), measurements from those photographs have been limited ({\it e.g.,} only selected umbral areas) or unavailable in machine-readable form.

Here we examine the RGO record for variations in the ratio of the penumbral-to-umbral areas in sunspot groups.
The RGO measurements of these quantities were made from photographic plates obtained at a small network of observatories using, for the most part, identical instruments.
Five Dallmeyer photoheliographs were constructed to make observations of the 1874 transit of Venus.
These instruments had four-inch aperture objective lenses with five-foot focal lengths.
The half-inch image was originally magnified to a four-inch image on photographic plates.
In September of 1875 the first of these photoheliographs replaced the Kew photoheliograph that had been in use at RGO since 1873.
In 1878 the second Dallmeyer photoheliograph was installed at Dehra D\^un in Northwest India.
In February of 1884 new magnifyers were used to project an eight-inch image.
In 1885 the third Dallmeyer photoheliograph was installed at the Royal Alfred Observatory in Mauritius.
With the addition of the third observatory, very few days went by without a photographic plate being taken (photographs were typically obtained on more than 360 days per year).
1885 also saw an effort to fill in previous missing days using photographs from observatories at Harvard College, Cambridge, Mass. USA and Melbourne, Australia.

The Dallmeyer photoheliograph at Greenwich was replaced by a Thompson photoheliograph with nine-inch aperture in 1898.
(The Dallmeyer at Greenwich still saw occasional use as the Thompson was often shipped off to observe eclipses.)
In 1904 the fourth Dallmeyer photoheliograph was installed at Kodaikanal in South India -- virtually assuring daily coverage.

In all of these instruments the Sun's visible image was focused on the photographic plates with a set diameter of 8 inches (7.5 inches for the Thompson).
The sunspot areas were measured off of these plates by first positioning a glass diaphragm etched with squares with sides of 0.01 inches in close contact to the photographic plate.
Two observers would then count squares and partial squares covering whole sunspots and sunspot umbrae.
The areas reported were the average of these two independent measurements and included both the projected areas as observed on the disk and areas corrected for foreshortening as a function of center-to-limb distance.

Identifying the umbral and penumbral boundaries (off of photographic plates with the human eye and brain) obviously depends to some degree on scattered light, seeing, telescope, and photographic emulsion.
Nonetheless, \inlinecite{Baranyi_etal01} have found that one-to-one comparisons of individual umbral and penumbral areas using modern photographs from Drebrecen, Rome, and the Soviet observatories reporting in the {\it Solnechnie Dannie} give very consistent results (at worst 5-15\% systematic differences for the smallest sunspots) even when very different telescopes, emulsions, and measuring methods were in use.

At the RGO and its associated observatories the same telescopes and photographic emulsions were in use with few exceptions ({\it e.g.} while RGO, Mauritius, and Kodaikanal were using dry gelatine plates starting in 1882, Dehra D\^un continued to use wet colloidion plates until 1902 and, while the Dallmeyer photoheliographs were in continuous use at Dehra D\^un, Mauritius, and Kodaikanal the larger Thompson photoheliograph came into use at the RGO in 1898).
Furthermore, the same measuring methods were used for all of the plates incorporated in the RGO database and the pair of observers measuring the plates often had a choice of plates from two or more sites to select the best data for the day.
These factors have made the RGO data the data of choice for long-term studies of sunspot areas.

\section{Penumbra-to-Umbra Area Ratio}

Early authors examined the ratio of penumbral-to-umbral radii for single regular sunspots, which is equivalent to
\begin{equation}
	q \equiv \sqrt{\frac{A_{\rm W}}{A_{\rm U}}},
\end{equation}
where $A_{\rm W}$ is the area of the whole sunspot or sunspot group, and $A_{\rm U}$ is the area of the umbra, both corrected for foreshortening based on the positions of sunspots between disk center and limb.
This made good sense when examining single regular sunspots, but is less sensible when including complex groups containing many irregular sunspots whose radii were poorly defined.

Here we chose to take the ratio of the penumbral area to the umbral area in the same way as in \inlinecite{Antalova71},
\begin{equation}
	 q' = \frac{A_{\rm W} - A_{\rm U}}{A_{\rm U}} = q^2 - 1.
\end{equation}
This is calculated for all 161,839 sunspot group records in the RGO {\it Photoheliographic Results}.

\begin{figure} 
\centerline{\includegraphics[width=0.8\textwidth]{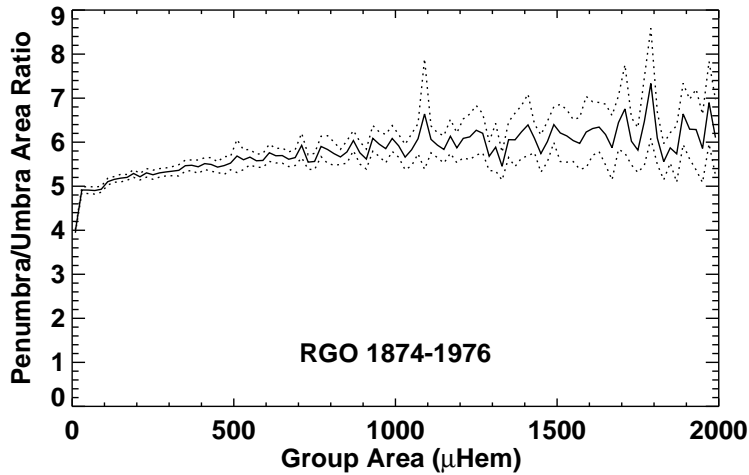}}
\caption{Penumbra-to-umbra area ratio as a function of total (umbra + penumbra) sunspot group area for all sunspot groups in the RGO database. Dotted lines represent $2\sigma$ errors. This ratio increases from 5 to 6 as the sunspot group areas increase from $100\ \mu$Hem to $2000\ \mu$Hem but with a sharp decrease to 4 for the smallest sunspot groups.}
\end{figure}

Figure 1 shows the average of this ratio as a function of the whole group area for all sunspot group observations in the database.
The data are binned according to their corrected whole sunspot area in bins $20\ \mu$Hem wide from 0 to $2000\ \mu$Hem.
This average is calculate by taking the ratio for each (daily) observation of each group and
averaging it with all other ratios for observations in the same area bin.

The area ratio increases from 5 to 6 as the sunspot group areas increase from $100\ \mu$Hem to $2000\ \mu$Hem, consistent with the results reported by \inlinecite{TandbergHanssen56} and \inlinecite{Antalova71} for all sunspot groups, but in contrast to the earlier work of \inlinecite{Waldmeier39B} and \inlinecite{Jensen_etal55} for single isolated sunspots.
This increasing ratio with increasing sunspot group area must be due to the production of additional penumbral area between sunspot umbrae in the larger, more complex, sunspot groups.

\section{Variations with Latitude}

We examined possible variations with latitude by calculating the average ratios for sunspot groups in four latitude ranges, $0^\circ$ to $10^\circ$, $10^\circ$ to $20^\circ$, $20^\circ$ to $30^\circ$, and $30^\circ$ to $50^\circ$, without regard to hemisphere.
The penumbra-to-umbra area ratios as functions of whole sunspot area are shown in Figure 2 for these four latitude ranges.
The results are consistent with no variation with latitude as was found by \inlinecite{Antalova71}.

\begin{figure} 
\centerline{\includegraphics[width=0.8\textwidth]{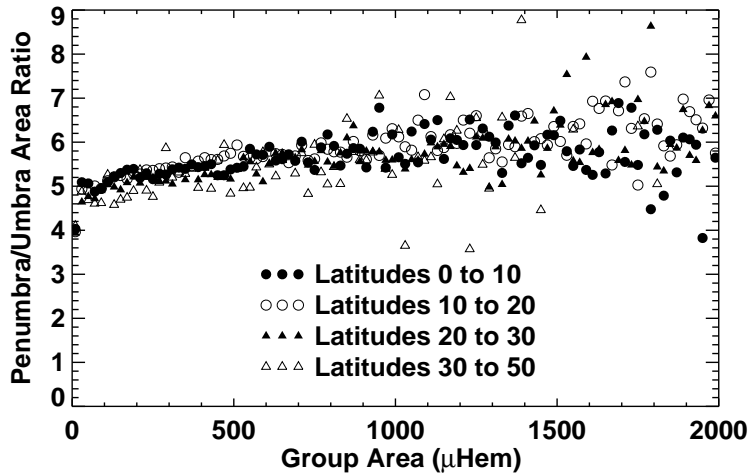}}
\caption{Penumbra-to-umbra area ratio as a function of total sunspot group area for sunspot groups in various latitude ranges. This relationship does not change with latitude.}
\end{figure}

\section{Variations with Sunspot Cycle Phase}

Variations with sunspot cycle phase were reported previously.
\inlinecite{Jensen_etal55} found a tendency for higher area ratios at cycle maxima than at minima, but in their follow-on study \cite{Jensen_etal56} noted that this tendency seems weaker or nonexistent in the later cycles.
\inlinecite{TandbergHanssen56} and \inlinecite{Antalova71} also reported variations with phase of the solar cycle with a more extended penumbral area at maxima.
\inlinecite{TandbergHanssen56} noted that the drop in $q$ for small sunspots was slightly different at cycle minima and maxima and that the scatter in $q$ was greater at cycle maxima.
However, his quantification of these characteristics showed substantial variation from one cycle to the next.
\inlinecite{Antalova71} found that, while $q'$ varied widely from one cycle minimum to the next, in general $q'$ was higher at cycle maximum than it was at the previous minimum.

Here we find that variations with the cycle phase are not apparent when data from different phases of the solar cycle are grouped together for all cycles in the RGO database.
Figure 3 shows the penumbra-to-umbra area ratio as a function of the total sunspot group area for four different phases of the sunspot cycle: minimum (from minimum minus three years to minimum plus one year), rising (from minimum plus two years to minimum plus three years), maximum (from minimum plus four years to minimum plus six years), declining (from minimum plus seven years to minimum plus eight years).

This lack of variation with cycle phase is consistent with the lack of variation with latitude since sunspots are found at lower latitudes as the cycle progresses.
It also suggests, as was noted by \cite{Jensen_etal56}, that the sunspot-cycle-dependent behavior seen in the earlier studies was primarily due to data from the earliest cycles.

\begin{figure} 
\centerline{\includegraphics[width=0.8\textwidth]{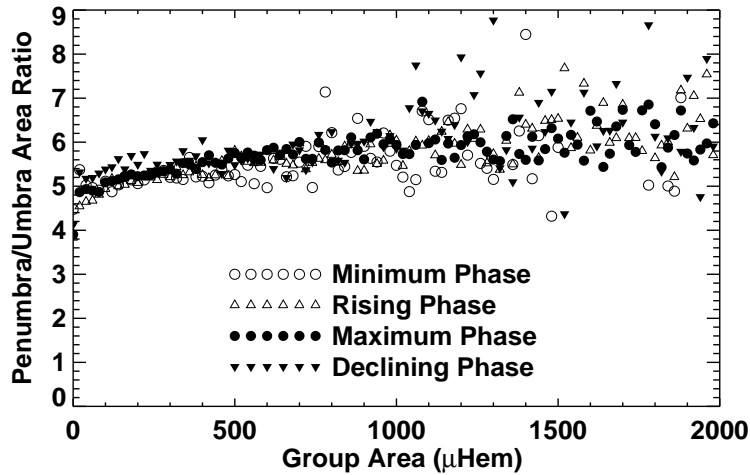}}
\caption{Penumbra-to-umbra area ratio as a function of total sunspot group area for sunspot groups at various phases of the sunspot cycle as indicated by the different symbols. This relationship does not change substantially with cycle phase.}
\end{figure}

\section{Secular Variations}

Examining the penumbral-to-umbral area ratio for individual sunspot cycles reveals a curious behavior.
The penumbral-to-umbral area ratio for the early cycles (cycles 12-14) and the late cycles (cycles 19 and 20) shows one type of behavior -- the small sunspot groups (group area $< 100\ \mu$Hem) show an increase in penumbral area (Figure 4).
The penumbral-to-umbral area ratio for the cycles in between (cycles 15-18) shows a different type of behavior -- the small sunspot groups (group area $< 100\ \mu$Hem) show a pronounced decrease in penumbral area (Figure 5).

\begin{figure} 
\centerline{\includegraphics[width=0.8\textwidth]{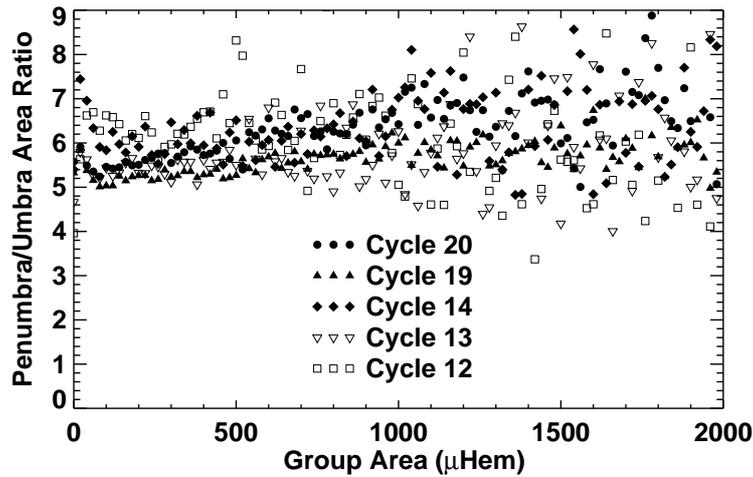}}
\caption{Penumbral-to-umbral area ratio as a function of total sunspot group area for sunspot groups in sunspot cycles 12-14 and 19-20 as indicated by the different symbols. These sunspot cycles exhibit an {\em increase} in penumbral areas for small sunspot groups.}
\end{figure}

\begin{figure} 
\centerline{\includegraphics[width=0.8\textwidth]{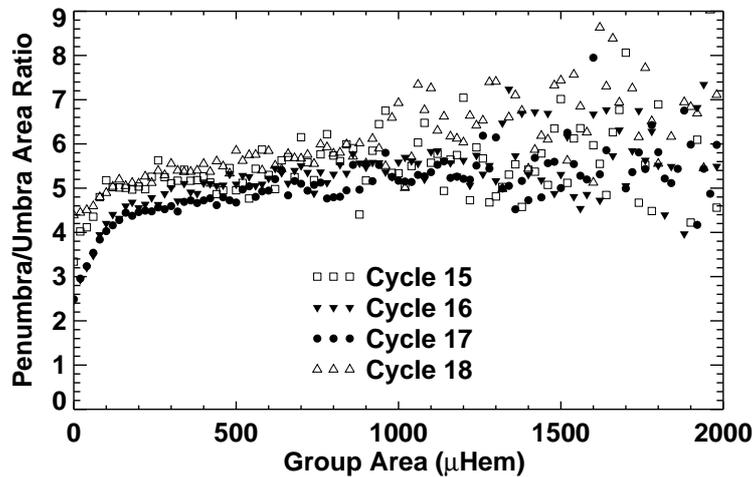}}
\caption{Penumbral-to-umbral area ratio as a function of total sunspot group area for sunspot groups in sunspot cycles 15-18 as indicated by the different symbols. These sunspot cycles exhibit a substantial {\em decrease} in penumbral areas for small sunspot groups.}
\end{figure}

The evolution of this behavior can be further examined by calculating the average penumbral-to-umbral area ratio on a yearly basis for sunspot groups with areas $> 100\ \mu$Hem and for sunspot groups with areas $< 100\ \mu$Hem.

Figure 6 shows the yearly averages of the penumbral-to-umbral area ratio for sunspot groups with areas $> 100\ \mu$Hem.
This ratio is noisy early on but nonetheless shows a weak trend from a value of about 6 in the early cycles to a value of about 5 in the ``in between'' cycles and then rising again to a value of about 6 in the late cycles.
The values plotted in this figure for years prior to 1955 correspond almost one-to-one with those plotted by Jensen, Nord\o, and Ringnes (1955, 1956) in Figure 1 of each paper for all sunspots (the bottom line in those figures).
However, the addition of statistical error bars in Figure 6 here indicates how uncertain the values were for the earliest cycles. 

\begin{figure} 
\centerline{\includegraphics[width=0.8\textwidth]{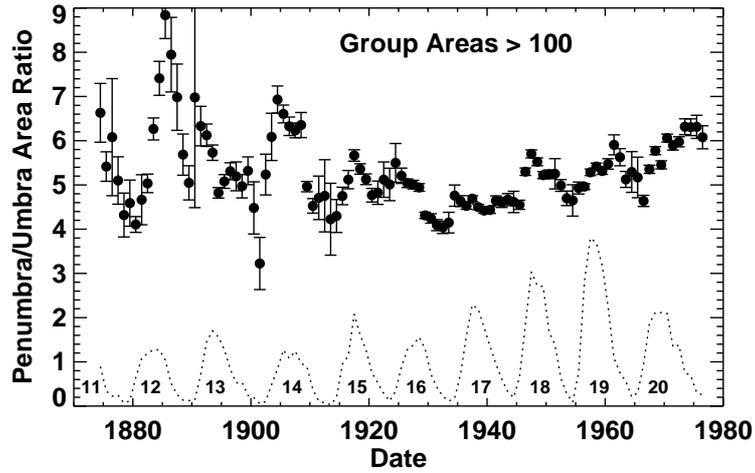}}
\caption{Yearly averages of the penumbral-to-umbral area ratio as a function of time for sunspot groups with areas $> 100\ \mu$Hem are shown by the filled circles with $2\sigma$ error bars. The yearly sunspot number divided by 50 is shown with the dotted line with sunspot cycle numbers for reference.}
\end{figure}

Figure 7 shows the yearly averages of the penumbral-to-umbral area ratio for sunspot groups with areas $< 100\ \mu$Hem.
Again this ratio is noisy early on but then shows a dramatic variation.
It reaches a high of $>7$ in 1905 and 1906, drops almost monotonically to a value $<3$ in the 1930s, and then rises almost monotonically back to a value $>7$ in 1961.

This multi-decadal variation of more than a factor of 2 does not appear to be tied to solar activity.
The sunspot cycle amplitudes were increasing almost monotonically from 1905 to 1960 while these penumbral areas first dropped by a factor of more than 2 and then increased again to former levels.

\begin{figure} 
\centerline{\includegraphics[width=0.8\textwidth]{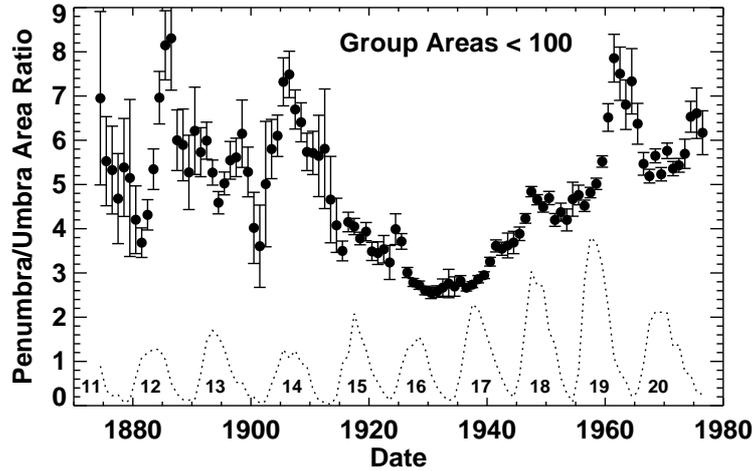}}
\caption{Yearly averages of the penumbral-to-umbral area ratio as a function of time for sunspot groups with areas $< 100\ \mu$Hem are shown by the filled circles with $2\sigma$ error bars. The yearly sunspot number divided by 50 is shown with the dotted line for reference.
From 1905 to 1960 the penumbral areas of these small sunspots decrease to less than half their former areas and then returned to previous levels.}
\end{figure}

\section{Conclusions}

The ratio of the sunspot group penumbral-to-umbral area for all sunspot groups is $\approx 5.5$, but with an increase from 5 to 6 as the group area increases from 100 to 2000 $\mu$Hem.
The average value is consistent with all earlier studies.
(Note that \inlinecite{Vaquero_etal05} recently found similar values from early observations with the Kew photoheliograph by de la Rue from the years 1862 to 1866.)
The increase with area is consistent with studies of all sunspot groups \cite{TandbergHanssen56, Antalova71}, but an opposite trend with area was found for individual isolated sunspots (Waldmeier, 1939; Jensen, Nord\o, and Ringnes, 1955, 1956).
This suggests, as was first noted by \inlinecite{TandbergHanssen56}, that the proximity of other sunspots in a group increases the penumbral area.
It is also consistent with MHD models of sunspot structure \cite{Rempel_etal09}.

Here we find that this behavior does not vary with sunspot group latitude or the phase of the sunspot cycle.
(It appears that the earlier reports of variations with sunspot cycle phase were heavily influenced by the earlier and noisier data from the RGO.) 

The key finding here is that there appears to be a curious and systematic variation in the penumbral areas with time over the 100-year record.
The larger sunspot groups (areas $> 100\ \mu$Hem) show only a slight decrease in relative penumbral area over the first 50 years, followed by a return to the original values over the last 50 years.
The smaller sunspot groups (areas $< 100\ \mu$Hem) show a dramatic decrease in relative penumbral area over the first 50 years, followed by a return to the original values over the last 50 years.
These variations appear to be gradual with little or no connection to the sunspot cycle itself and without any discontinuities that might indicate changes in equipment or observers.
This 100-year variation has a time scale similar to the Gleissberg cycle for cycle amplitudes \cite{Gleissberg39, Hathaway10}.
This curious behavior could have consequences for other aspects of solar variability -- irradiance variability in particular.
Small sunspot groups vastly outnumber large sunspot groups.
\inlinecite{Bogdan_etal88} found that sunspot umbral areas have a log-normal distribution with $10\ \mu$Hem umbrae outnumbering $100\ \mu$Hem umbrae by a factor of $10^3$ or more.
If the penumbrae of these small sunspots vary in area by a factor of more than two over long (multi-decadal) time-scales, then estimates of the the Sun's total irradiance based on sunspot number or sunspot area could be substantially in error.

This behavior could have implications for models of penumbra formation.
While the difference in penumbral area for large single sunspots and large groups with the same total area clearly indicates the importance of nearby sunspots on the formation of penumbrae, it is unclear which mechanism might be involved in producing this long, slow variation in the penumbral areas of the smallest sunspots groups (which usually consist of single sunspots).
There are indications of a century-long increase in the solar magnetic flux open to the solar wind \cite{Lockwood_etal99}.
Could the global field of the Sun change the penumbrae of the small sunspots?

This behavior could also have implications for the solar dynamo.
Just as the report by \inlinecite{PennLivingston06} on recent changes in average magnetic field strength and emergent intensity in sunspot umbrae suggests different dynamo behavior, this curious behavior of the small sunspot penumbrae may also suggest a change in dynamo behavior.

Of course, this curious behavior could be an observational artifact -- a product of changing observing methods and/or observers.
E. Walter Maunder was responsible from 1874 to his retirement in 1919.
The penumbral-to-umbral area ratio for the small sunspot groups dropped smoothly from 1905 to 1919, but then continued to decline to its low point in 1930.
This smooth variation does not have any step-like changes that would be expected from changes in observing methods and/or observer.

Confirmation from other observatory records would of course be extremely helpful.
Both Mt. Wilson and Kodaikanal have photographic plate collections that encompass this time period of curious behavior.
While the Ca{\sc ii} K-line photographs from Mt. Wilson have been digitized, the white-light image digitization has not yet been completed.
Very recently \inlinecite{Ravindra_etal13} announced the completion of the initial digitization of the Kodaikanal white-light photographs.
Currently, only low-resolution ($800 \times 800$) uncalibrated images are available.
Hopefully, these will be superseded by high resolution ($4k \times 4k$) calibrated images soon.

%%%%%%%%%%%%%%%%%%%%%%%%%%%%%%%%%%%%%%%%%%%%%%%%%%%%%%%%%%%%%%%%%%%%%%%%%%%
%% Appendix
%
% \appendix   

%%%%%%%%%%%%%%%%%%%%%%%%%%%%%%%%%%%%%%%%%%%%%%%%%%%%%%%%%%%%%%%%%%%%%%%%%%%
%% Acknowledgements
%
 \begin{acks}
The author thanks Lisa Upton for reviewing the paper and NASA for its support of this research through a grant from the Heliophysics Causes and Consequences of the Minimum of Solar Cycle 23/24 Program to NASA Marshall Space Flight Center.
Comments by an anonymous referee led to significant improvements in the paper.
Most importantly, the author thanks the American taxpayers for supporting scientific research in general and this research in particular.
 \end{acks}

%%% %%%%%%%%%%%%%%%%%%%%%%%%%%%%%%%%%%%%%%%%%%%%%%%%%%%%%%%%%%%
%% Bibliography
%

\end{article} 

\begin{thebibliography}{}
 \bibitem[\protect\citeauthoryear{Antalov\'a}{1971}]{Antalova71}
   Antalov\'a, A.: 1971, {\it Bull. Astro. Inst. Czechoslovakia} {\bf 22}, 352.

 \bibitem[\protect\citeauthoryear{Baranyi {\it et al.}}{2001}]{Baranyi_etal01}
   Baranyi, T., Gy\"ori, L., Ludm\'any, A., Coffey, H. E.: 2001,
   {\it Mon. Not. Roy. Astron. Soc.} {\bf 323}, 223.

 \bibitem[\protect\citeauthoryear{Bogdan {\it et al.}}{1988}]{Bogdan_etal88}
   Bogdan, T. J., Gilman, P. A., Lerche, I., Howard, R.: 1988,
   {\it Astrophys. J.} {\bf 327}, 451.

 \bibitem[\protect\citeauthoryear{Borrero and Ichimoto}{2011}]{BorreroIchimoto11}
   Borrero, J. M., Ichimoto, K.: 2011,
   {\it Living Rev. Solar Phys.} {\bf 8}, (4), \newline
   http://solarphysics.livingreviews.org/Articles/lrsp-2011-4/

 \bibitem[\protect\citeauthoryear{Foukal and Lean}{1990}]{FoukalLean90}
   Foukal, P., Lean, J.: 1990, {\it Science} {\bf 247}, 556.

 \bibitem[\protect\citeauthoryear{Gleissberg}{1939}]{Gleissberg39}
   Gleissberg, W.: 1939, {\it Observatory} {\bf 62}, 158.

 \bibitem[\protect\citeauthoryear{Hale}{1908}]{Hale08}
   Hale, G. E.: 1908, {\it Astrophys. J.} {\bf 28}, 315.

 \bibitem[\protect\citeauthoryear{Hathaway}{2010}]{Hathaway10}
   Hathaway, D. H.: 2010,
   {\it Living Rev. Solar Phys.} {\bf 7}, (1), \newline
   http://solarphysics.livingreviews.org/Articles/lrsp-2010-1/

 \bibitem[\protect\citeauthoryear{Jensen, Nord\o, and Ringnes}{1955}]{Jensen_etal55}
   Jensen, E., Nord\o, J., Ringnes, T. S.: 1955, {\it Astrophys. Norvegica} {\bf 5}, 167.

 \bibitem[\protect\citeauthoryear{Jensen, Nord\o, and Ringnes}{1956}]{Jensen_etal56}
   Jensen, E., Nord\o, J., Ringnes, T. S.: 1956, {\it Ann. Astrophys.} {\bf 19}, 165.

 \bibitem[\protect\citeauthoryear{Lockwood, Stamper, and Wild}{1999}]{Lockwood_etal99}
   Lockwood, M., Stamper, R., Wild, M. N.: 1999, {\it Nature} {\bf 399}, 437.

 \bibitem[\protect\citeauthoryear{Nicholson}{1933}]{Nicholson33}
   Nicholson, S. B.: 1933, {\it Publ. Astron. Soc. Pacific} {\bf 45}, 51.

 \bibitem[\protect\citeauthoryear{Penn and Livingston}{2006}]{PennLivingston06}
   Penn, M. J., Livingston, W.: 2006, {\it Astrophys. J. Lett.} {\bf 649}, L45.

 \bibitem[\protect\citeauthoryear{Ravindra {\it et al.}}{2013}]{Ravindra_etal13}
   Ravindra, B., Priya, T. G., Amareswari, K., Priyal, M., Nazia, A. A.,
   Banerjee, D.: 2013, {\it Astron. Astrophys.} {\bf 550}, A19.

  \bibitem[\protect\citeauthoryear{Rempel and Schlichenmaier}{2011}]{RempelSchlichenmaier11}
   Rempel, M., Schlichenmaier, R.: 2011,
   {\it Living Rev. Solar Phys.} {\bf 8}, (3), \newline
   http://solarphysics.livingreviews.org/Articles/lrsp-2011-3/
   
  \bibitem[\protect\citeauthoryear{Rempel {\it et al.}}{2009}]{Rempel_etal09}
   Rempel, M., Sch\"ussler, M., Cameron, R. H., Kn\"olker, M. : 2009,
   {\it Science} {\bf 325}, 171.

 \bibitem[\protect\citeauthoryear{Tandberg-Hanssen}{1956}]{TandbergHanssen56}
   Tandberg-Hanssen, E.: 1956, {\it Astrophys. Norvegica} {\bf 5}, 207.

 \bibitem[\protect\citeauthoryear{Vaquero {\it et al.}}{2005}]{Vaquero_etal05}
   Vaquero, J. M., Gordillo, A., Gallego, M. C., S\'anchez-Bajo, F.,
   Garcia, J. A.: 2005, {\it Observatory} {\bf 125}, 152.

 \bibitem[\protect\citeauthoryear{Waldmeier}{1939}]{Waldmeier39B}
   Waldmeier, M.: 1939, {\it Astron. Mitt. Z\"urich} {\bf 14} (138), 439. 

 \end{thebibliography}
\end{document}